# Identifying influential patents in citation networks using enhanced VoteRank centrality

Freitas, João C.S. *, Barbastefano, Rafael **, Carvalho, Diego **

* FGV – Getulio Vargas Foundation. Praia de Botafogo, 190 – Botafogo – Rio de Janeiro/RJ - Brazil

joaocfreitas@gmail.com

** Cefet – Federal Centre for Technological Education Celso Suckow da Fonseca. Av. Maracanã, 229 - Maracanã – Rio de Janeiro/RJ – Brazil

barbastefano@gmail.com

d.carvalho@ieee.org

## Abstract

This study proposes the usage of a method called VoteRank, created by Zhang et al. (2016), to identify influential nodes on patent citation networks. In addition, it proposes enhanced VoteRank algorithms, extending the Zhang et al. work. These novel algorithms comprise a reduction on the voting ability of the nodes affected by a chosen spreader if the nodes are distant from the spreader. One method uses a reduction factor that is linear regarding the distance from the spreader, which we called VoteRank-LRed. The other method uses a reduction factor that is exponential concerning the distance from the spreader, which we called VoteRank-XRed. By applying the methods on a citation network, we were able to demonstrate that VoteRank-LRed improved performance in choosing influence spreaders more efficiently than the original VoteRank on the tested citation network.

# 1 - Introduction

Patent citations are often used as a direct means for distinguishing technologies among themselves, but the amount of times a patent is cited is also an indicative of its value (Jang, Lo, & Chang, 2009). Moreover, a patent citation usually represents knowledge contained in one patent and extended by a newer one, meaning a knowledge flow (Jaffe, Trajtenberg, & Henderson, 1993). Patent citations have also been used to detect technology diffusion and forecasting (Chang, Lai, & Chang, 2009) or knowledge spillover (Xiang, Cai, Lam, & Pei, 2013).

Since citations establish relationships between patents (actors), social networks theory can be applied to build patent citations networks and acquire better comprehension about citations (Chang, Lai, & Chang, 2009). Therefore, many studies have been made using patent citation networks, such as (Hung & Wang, 2010) or (Kim, Cho, & Kim, 2014). Xiang et al. (2013) applied a spillover technique to increase the amount of relationships among patents in citation networks. The spillover technique creates an arc in the network from an older to a newer patent if they share at least one common author, even if there was no citation between them on the original network. The goal is to explicit the relationship between these patents, because tacit knowledge acquired during the first work was implicitly used by the common author on the newer patent.

Centrality is a key concept in social networks. The most important or prominent actors of a network are said to be central (Wasserman & Faust, 2006). A node centrality can be calculated in many different ways, the most usual being degree, closeness, and betweenness, which can be found on most social network analysis books, such as (Wasserman & Faust, 2006) or (De Nooy, Mrvar, & Batagelj, 2005). However, increasingly more ways of expressing the centrality of nodes, like LocalRank, ClusterRank, PageRank, H-Index and others, have been appearing lately, as gathered by Lü et al. (2016).

Among these newer methods, lies the VoteRank algorithm (Zhang, Chen, Dong, & Zhao, 2016), known for its swiftness and efficiency and for providing several spreaders simultaneously. In a patent citation network, a spreader can be understood as a document that is cited throughout an innovation chain. Since the efficient detection of important members from a network is a core need for the understanding of knowledge flow through citations or patents values, this study discloses this new method to the field of patent citation network analysis. It proposes some modifications on VoteRank in order to find better spreaders.

This study is structures as follows: section 2 presents the Centrality, VoteRank, and SIR criterion; section 3 discloses two variations of VoteRank; and section 4 displays the achieved results.

# 2 – Centrality

Centrality is a key concept in social network analysis. One of most important applications of graph theory in SNA is the identification of prominent vertices (actors) in a network (Wasserman & Faust, 2006). Networks are abstract concepts; thus, the definition of centers becomes difficult, since there is no similarity with the physical concept of centroid.

The first centrality concept is related to the number of ties connected to a vertex. A central actor with several connections is involved in more relations than an actor with fewer connections. This definition is both topological and local. Other traditional centrality measures are related to the average distance from one vertex to the others (closeness), or to network information flows (betweenness).

The relative importance of each actor within the network can be recognized through centrality metrics. In the scope of this study, the following centrality measures were employed: degree, closeness, and betweenness. The so-called giant component is yet another interesting variable to observe in a citation network. The following are the centrality metrics and the giant component definitions as seen on (Wasserman & Faust, 2006) and (De Nooy, Mrvar, & Batagelj, 2005).

Let $G=\{V, E\}$ be a undirected graph representing a network, where $V$ is the set of nodes and $E$ is the set of edges in the network. The degree centrality $Dc$ of a node is measured by the number of nodes directly connected to the node $v$: $Dc(v) = degree(v)$. On directed networks, we can also have the in-degree which is defined by the number of incoming arcs on the node, and also the out-degree, defined by the number of outgoing arcs from the node.

The closeness centrality *Cc* of a node *v* is defined by the inverse of the average of all shortest paths (geodesic paths) from *v* to all other *n* nodes on the network v: $Cc(v) = \frac{n-1}{\sum_y d(v,y)}$. The shorter the sum of distances from *v* to other nodes is, the higher the closeness centrality gets. Unlike the degree centrality, which is local, the closeness centrality is a global measure, since it is defined by the entire network topology.

The betweenness centrality *Bc* of a node *v* indicates the percentage of information flows on the network that goes through *v*. It is defined as a proportion of the amount *T(x,v,y)* of shortest paths between every pair of nodes in the network where *v* lies in the path and the amount *T(x,y)* of all shortest paths in the network: $Bc(v) = \sum_{x,y \neq v} \frac{T(x,v,y)}{T(x,y)}$. The betweenness centrality is measurable only on directed networks since the direction of the flow is relevant. It is also a global measure, like the closeness centrality.

On citation or co-authoring networks, it is common to verify that a node is not connected to every other node in the network. On these cases, most analysis happens on a so-called giant component that comprises many nodes of the network. The giant component can be defined as the largest connected component of a network and normally it contains a significant fraction of the nodes.

## 2.1 - VoteRank

The VoteRank algorithm is a method to calculate the centrality of nodes on directed or undirected networks. In this context, the centrality itself reflects the node's capacity for spreading news or influence over a network. The VoteRank algorithm provides a better choice of influential nodes than other comparable methods, such as PageRank or Degree, as demonstrated by Zhang, Chen, Dong & Zhao (2016).

The VoteRank algorithm creates a concept of votes between connected nodes. Each node can be voted by its immediate neighbors, each of these neighbors having its own voting ability. The voting score of a node is the sum of the voting ability of its neighbors. A node with high voting score is more relevant than a node with a low voting score.

The VoteRank algorithm begins with every node having a voting ability of 1. On each turn, the VoteRank algorithm computes the voting score for every node in the network, and picks the node with the highest voting score as an influence spreader. On undirected networks, the voting score is the sum of all neighbors' voting ability within a distance to the spreader equal to one. On directed networks, the voting score is the sum of the voting abilities only from the neighbors that receive information from the spreader, which Zhang et al. refer to as out-neighbors.

Once a spreader is chosen, its voting ability and voting score are zeroed, and the voting ability of each of its neighbors is reduced by a factor of $\frac{1}{<k>}$, where <k> is the average degree of the network, or the average out-degree on directed networks. If there are more spreaders to choose from the network, the VoteRank algorithm once again computes the voting scores for the network's nodes, excluding the spreaders that had already been chosen, and picks the current node with the highest voting score as a spreader. Once more, the spreader's score and voting ability are zeroed, and the voting ability of its immediate neighbors is decreased. The method repeats this cycle until a satisfying number of spreaders is chosen.

The VoteRank method is computationally efficient. On each iteration, the method reduces the voting ability values for a small set of nodes, and then computes the new voting scores for their neighbors alone. The method is quite swift, since it only operates on a small section of the network at a time. Moreover, the VoteRank algorithm provides an interactive form for choosing important nodes on the network, and it is more efficient than other methods explored in the study, according to the SIR spreading model.

## 2.2 – SIR

The methods comparison on the study by Zhang et al. was performed using the SIR spreading model. The SIR spreading model is a *de facto* standard used for comparing algorithms that search for the most important or central nodes on networks. It has been extensively utilized on several studies such as (Wang, Zhang, Zhao, & Yi, 2016), (Lü, Zhou, Zhang, & Stanley, 2016), (Liu, Tang, Zhou, & Do, 2015) or (Malliaros, Rossi, & Vazirgiannis, 2016), among others. The following information in this section is based on the study of Pastor-Satorras et al. (2015).

The standard susceptible-infected-recovered (SIR) epidemic model is applied to simulate how fast and extensively an infection can be spread over a network.

On this model, all nodes on the network are marked as susceptible, infected or recovered. A susceptible node is a node that can be infected. An infected node is a node that can spread the infection to susceptible nodes. A recovered node is a node that once was but no longer is infected, and became immune to infection. There are variations, e.g. the recovered can be infected again (SIS), but the SIR spreading model is more commonly used.

The method starts with a set of one or more spreaders being defined as infected. All other nodes on the network are marked as susceptible. On each turn the infected nodes try to spread their infection to neighbors with a success probability μ. After the spreading, and still on the same turn, the newly infected nodes can become no longer infected with probability β and, in case of success, get marked as recovered. This ends the current turn, and a new turn begins, with new infections occurring and infected nodes being recovered. The simulation ends when all nodes are either susceptible or recovered.

In our scenario, the infection is the spreading of information. Clearly, the spreaders are the nodes chosen with VoteRank or any other method. Simulations using several ratios between μ and β can be made, determining how fast the spreading occurs and/or how extensive the spreading is. This ratio is usually expressed as λ ($\lambda = \frac{\mu}{\beta}$). If λ>0, the number of infected nodes grow exponentially, so simulations usually have λ<0.

The methods are comparable in terms of how long the simulation took to end, or how fast the spreading grew. But the most important factor is how extensive was the spreading, in terms of the percentage of the network size that was affected, e.g. how many people received information broadcasted by a set of spreaders. It can be expressed by the function $R(t) = \frac{r}{n}$ , where *r* is the number of recovered nodes and *n* is the number of nodes on the network. When t = $t_{end}$, the simulation is over and there are no infected nodes left.

Generally, a method is considered to be more efficient than others if it determines a set of spreaders that affects a more significant portion of the network than the set computed by the other methods, provided the sets have a same size and the simulation parameters β and μ are equivalent. In a nutshell, the methods compete among themselves for the best choice of spreaders, and the SIR method judges among them.

## 3 - Alternative VoteRank algorithms

The VoteRank algorithm, as it was originally proposed, assumes that only the nearest neighbors have their voting ability reduced after a spreader has been chosen. On directed acyclic graphs, such as regular citation networks, even fewer nodes are affected, because only nodes connected with arcs leaving the spreader are computed (out-neighbors). This feature implies that a node on the network only spreads influence on its immediate neighborhood, via direct communication, arcs or edges.

However, we consider that a node spreads influence to a larger area of the network, both indirectly and through its closest neighbors. Assume that any given node A broadcasts information on the network. If information broadcasted by A reaches another node B being at a distance greater than one from node A, that is because of the spreading influence of node A anyway. Therefore, it is hypothesized that the voting ability for node B could also be affected by A, as B was influenced by A. Thus, the VoteRank algorithm can be somewhat extended.

We hereby propose an alternative VoteRank algorithm, where more neighbors can be affected by a chosen spreader node A. In theory, if there is a path between nodes A and B on the network, B could also have its voting ability reduced, regardless of how far apart they are.

The original VoteRank algorithm applies a reduction on the voting ability of its direct neighbors at a factor of $\frac{1}{<k>}$, where <k> is the average degree of the network, or the average out-degree on directed networks. We hypothesize that this reduction will also affect other, more distant neighbors, and that this reduction will depend on the distance from the spreader. The farther a node is from a spreader, the less its voting ability will be affected.

The first proposed alternative, called VoteRank-LRed, is applying a linear reduction on the voting ability at a factor of $\frac{1}{<k>d}$, where <k> is the average degree of the network, or the average out-degree on directed networks, and *d* is the distance from the node to the spreader. It is straightforward to verify that a node that is very far from the spreader nearly isn't affected by it.

The second proposed alternative, called VoteRank-XRed, is applying an exponential reduction on the voting ability $\frac{1}{<k>^d}$ where, again, <k> is the average degree of the network, or the average out-degree on directed networks, and *d* is the distance from the node to the spreader. This variation applies the reduction at a much faster pace, and the spreader's impact on distant neighbors is smaller than on VoteRank-LRed, though higher than on the original VoteRank.

Since the reduction factor decays with distance, the algorithm applies the reduction on the voting ability only up to a maximum distance from the spreader, in order to avoid excessive calculation. We decided to apply the reduction factor on nodes with a distance of at most <k> from the spreader, where <k> is the average degree node of the network, or the average out-degree on directed networks. On the next section, we will apply VoteRank, VoteRank-LRed, and VoteRank-XRed on some patent citation networks and compare their performances using the SIR model.

## 4 - Results

In order to apply and compare the VoteRank algorithms, we selected one real patent citation network created by the authors on previous study.
- 3DPRN describes citations amongst patents regarding 3D printing technology; it was obtained from Derwent World Patent Index on Aug-2014, and contains patents published until 2013.

Table 1 summarizes the main topological features of the network.

| Features | 3DPRN |
| --- | --- |
| nodes | 653 |
| arcs | 1416 |
| density | 0,003321 |
| average degree | 4,336907 |
| average out-degree | 2,168453 |
| nodes on giant component | 442 |
| % nodes on giant component | 67,7% |

**Table 1 – topological features of the test network**

We tested the network with the Degree, Closeness, Betwenness, VoteRank, VoteRank-LRed and VoteRank-XRed methods. For every test, we selected an arbitrary number of spreader nodes expressed in terms of a percentage *p* of the number of nodes on the network. The tests were performed with *p* values ranging from 0.0001 to 0.04. Many SIR simulations, with varying μ and β parameters, were performed for each network, method and *p* amount of spreaders. For each choice of *p*, μ and β, 1.000 independent simulations were performed. Therefore, the following graphs express average values from 1.000 simulations when *p,* μ and β were defined, or even more simulations when these parameters were not fully detailed. On this study, we performed over 3 million SIR simulations overall. Figure 1 shows the function R(t) for the percentage of nodes that recovered from the infection during the simulation.

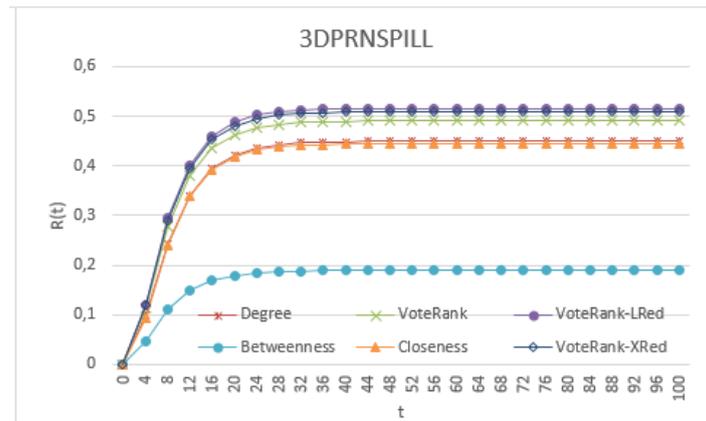

**Figure 1 – Extent of infection on networks 3DPRN where $p$=0.02, $\mu$=0.3 and $\beta$ =0.2**

Figures 1 shows that VoteRank-LRed had the best performance under these conditions.

Figure 2 shows how $R(t_{end})$ varies according to different values of $p$, given $\mu$ and $\beta$ are fixed.

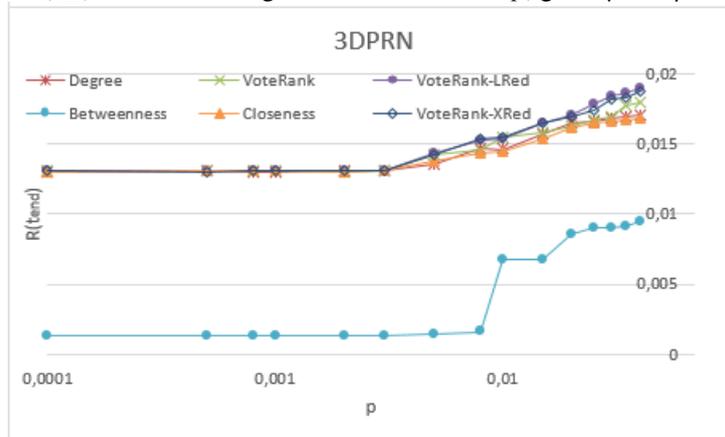

**Figure 2 – Variation of infection extent due to amount of spreaders chosen, given $\mu$ =0.4 and $\beta$ =0.2**

Results were consistent, across several values of $p$, that Betweenness was the weakest method, and the strongest were VoteRank-LRed, VoteRank-XRed, and VoteRank, in this order. On most simulations, results were inconclusive when the number of spreaders was smaller than 15, varying a lot and pending to Closeness as the method of choice. Above this threshold, VoteRank-LRed had the best performance, and VoteRank-XRed came second between the leading method and VoteRank.

So far, we have only observed the behavior of the methods in the variation of time and number of spreaders. Now we show in Figure 3 the comparison between the methods when the factor λ is changed. We fixed $p$=0.01 and $\mu$=0.5. Changing values of β, we had the following results. We show only the VoteRank methods on this figure for better visibility due to the scale of the graphs. As expected, as λ grows, more nodes are infected. The three methods behave equally.

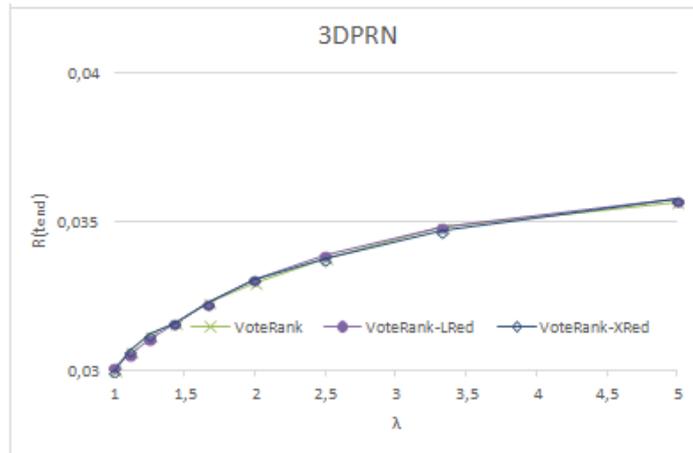
**Figure 3** – Variation of infection extent due to changes in SIR parameter β, with *p*=0.01 and μ =0.5.

Table 2 shows the distribution of victories of each method along with all the tests for each value of *p* and for each network. We created 66 possible combinations of values μ and β for each value of *p*. Again, it is clear that networks with smaller amounts of nodes were more affected by smaller values of *p*. As *p* increased, the usefulness of VoteRank-LRed increased as well. The EOLIC network is the largest one tested, and with *p* equal to or larger than 0.001, i.e. 15 spreaders, results consistently point to VoteRank-LRed as the best choice.

| p | 3DPRN ||||| 
|---|---|---|---|---|---|
|   | Closeness | Degree | VoteRank | VoteRank_LRed | VoteRank_XRed |
| 0.0001 | 7 | 12 | 25 | 13 | 9 |
| 0.0005 | 6 | 13 | 28 | 12 | 7 |
| 0.0008 | 13 | 4 | 23 | 20 | 6 |
| 0.001 | 8 | 2 | 29 | 14 | 13 |
| 0.002 | 8 | 8 | 27 | 17 | 6 |
| 0.003 | 5 | 10 | 29 | 12 | 10 |
| 0.005 |  |  | 41 | 18 | 7 |
| 0.008 |  |  |  | 3 | 63 |
| 0.01 |  |  | 43 | 13 | 10 |
| 0.015 |  |  |  | 52 | 14 |
| 0.02 |  |  |  | 66 |  |
| 0.025 |  |  |  | 66 |  |
| 0.03 |  |  |  | 66 |  |
| 0.035 |  |  |  | 66 |  |
| 0.04 |  |  |  | 66 |  |
| Total | 47 | 49 | 245 | 485 | 145 |

**Table 2** – Distribution of victories of each method through all tests.

VoteRank-LRed and VoteRank-XRed are somewhat more computationally intensive than original VoteRank. In the step where voting abilities are reduced, more nodes are searched and updated. The procedures needed to reach these other nodes up to distance *<k>* must be performed, and are not present on the original method. However, using good tree implementations and caching techniques, we can overcome the difference with no major impact. Moreover, it is a small price to pay if we are interested in achieving better results.

## 5 - Conclusion

The obtained results revealed that VoteRank-LRed had better results than original VoteRank in most scenarios. The variation hereby proposed is useful, and the hypothesis that the influence spread by the nodes on larger areas of the network should be taken in account has proven to be valid.

Further investigations can be conducted regarding an extension of VoteRank-LRed. This version of the method reduces the voting ability of nodes up to a $<k>$ distance from the spreader, where $<k>$ is the average degree of the network. The $<k>$ value was rather arbitrary, so simulations could be run to try to determine if there are better approximations for a distance up to which the neighbors should be affected. Other variations of the method could be tried, with other types of reductions instead of a linear one. Lastly, the voting process for choosing a spreader by calculating its score considers only its closest neighbors, i.e., those with a distance equal to one. Other versions of the method could be applied having the score take into account other neighbors, as we did with the reduction of the voting ability, perhaps up to a $<k>$ distance as well.

We are of the opinion that the application of these new methods on patent citation networks can bring new insights regarding the discovery of the most relevant patents in a network, and also help future analysis of the values and influences of both patents and scientists.